\documentclass[aps,showpacs,pra,preprint,amsmath,amssymb,superscriptaddress]{revtex4-1}
\usepackage{bm}
\usepackage{amssymb}
\usepackage{colordvi}
\usepackage{graphicx}
\usepackage{color}
\usepackage{hyperref}

\newcommand{\ov}[1]{\overline{{#1}}}
\newcommand{\be}{\begin{equation}}
\newcommand{\ee}{\end{equation}}
\newcommand{\bea}{\begin{eqnarray}}
\newcommand{\eea}{\end{eqnarray}}
\newcommand{\ep}{\epsilon}
\newcommand{\vep}{\varepsilon}

\newcommand{\ome}{\omega}
\newcommand{\Ome}{\Omega}

\def\Im {\mbox{Im}}

\begin{document}

\title{Photonic Architectures for Equilibrium High-Temperature
  Bose-Einstein Condensation in Dichalcogenide Monolayers}

\author{Jian-Hua Jiang}\email{Correspondence to jianhua.jiang.phys@gmail.com}
\affiliation{Department of Physics, University of Toronto, Toronto,
  Ontario, M5S 1A7 Canada}
\author{Sajeev John}
\affiliation{Department of Physics, University of Toronto, Toronto,
  Ontario, M5S 1A7 Canada}

\date{\today}
\maketitle

{\bf Semiconductor-microcavity polaritons are composite quasiparticles
  of excitons and photons, emerging in the strong coupling
regime. As quantum superpositions of matter and light, polaritons have 
much stronger interparticle interactions compared with
photons, enabling rapid equilibration and 
Bose-Einstein condensation (BEC). Current realizations 
based on 1D photonic structures, such as Fabry-P\'erot
microcavities, have limited light-trapping ability resulting in
picosecond polariton lifetime. We demonstrate, theoretically,
above-room-temperature (up to 590~K) BEC of long-lived polaritons in
MoSe$_2$ monolayers sandwiched by simple TiO$_2$ based 3D photonic
band gap (PBG) materials. The 3D PBG induces very
strong coupling of 40~meV (Rabi splitting of 62~meV) for as few
as three dichalcogenide monolayers. 
Strong light-trapping in the 3D PBG
enables the long-lived polariton superfluid to be robust against
fabrication-induced disorder and exciton line-broadening.}


Semiconductor microcavities with engineered photonic states are
a valuable platform to observe fundamental and emergent quantum
electrodynamic phenomena\cite{mc}. Polaritons are formed as a quantum
superposition of semiconductor-excitons and microcavity-photons when
the exciton-photon interaction is much larger than their decay rates. 
The ability to tailor the photonic modes, light-matter interaction
and polariton lifetime in semiconductor microcavities enables
versatile control of polaritons and opens the possibility of novel
quantum effects such as Bose-Einstein condensation (BEC) at elevated
temperatures\cite{bec1,bec2,bec3,bec4,deng}. 
{Picosecond time-scale quasiequilibrium} room-temperature
polariton BEC has been claimed in ZnO\cite{zno,zno2}, 
GaN\cite{gan,gan2,gan3}, and polymers\cite{polymer} in Fabry-P\'erot
microcavities. {This is possible when the thermalization time is
  even shorter than the polariton lifetime. However, for both
  scientific studies and applications, it is more favorable to achieve
  room-temperature polariton BEC with polariton lifetimes on the scale
  of 100~ps - 1~ns. Realization of long-lived, room-temperature,
  equilibrium polariton BEC remains an important target for
  fundamental research and practical application\cite{rt}.}


In this article, we demonstrate theoretically a route to
simultaneously achieve very strong exciton-photon coupling and long
polariton lifetime using a simple TiO$_2$ based photonic band gap
(PBG)\cite{pbg,pbg2} material sandwiching a planar quantum-well slab containing as
few as three monolayers of the transition-metal dichalcogenide
MoSe$_2$\cite{shen}. This architecture provides a realistic, technologically
accessible route toward equilibrium polariton BEC above
room temperature. In contrast, the
corresponding Fabry-P\'erot microcavity has picosecond polariton
lifetime due to radiative recombination into unwanted leaky (air)
modes that are degenerate with the microcavity mode.  These leaky modes
couple strongly to any finite area polariton condensate in the
quantum-well region. They are eliminated by replacing the
1D periodic structure with a 3D PBG material. Recently
WSe$_2$ and MoS$_2$ monolayers in 2D
photonic crystals have been experimentally realized\cite{2dphc1,2dphc2}. However, a
complete 3D PBG is absent in those cavities and radiative decay
of polaritons is not suppressed effectively. Another
realization of a MoS$_2$ monolayer in a Fabry-P\'erot
microcavity\cite{mos2-fp} reveals exciton-photon coupling of 31.5~meV
but very rapid (sub-picosecond) polariton decay and significant
line-broadening of 38.5~meV.

In a MoSe$_2$ monolayer the exciton binding energy is as large as 0.55~eV
and the exciton Bohr radius is as small as 1.2~nm\cite{shen,mose2}.
Theoretical calculation\cite{shen} predicts that 6\% of incident light is absorbed
during transmission through a MoSe$_2$ layer\cite{shen}, indicating
very strong light-matter interaction. As we show, this enables
high-temperature BEC with only three quantum-well monolayers of
MoSe$_2$. Previous considerations of high-temperature, equilibrium BEC
have involved on the order of 100 quantum-wells\cite{cdte}.
Another special property of MoSe$_2$ is that due to correlation
between wave-vector valley and spin polarizations at the electronic
band edge of MoSe$_2$, left- and right-circularly polarized light are
coupled to different exciton spin and valley states\cite{yao}. In
other words, optical, valley, and spin polarizations are correlated
with each other. Information encoded in the spin and valley channels
of the condensate can be uncovered by measuring the polarization of
emitted light. This may provide a new degree of freedom to store and
manipulate coherent quantum information in the predicted
polariton condensate\cite{electric,gating,gating2}.

\section*{Optical Microcavity Architectures}

A fundamental challenge to the realization of above-room-temperature
BEC is the accurate fabrication of complex photonic
nanostructures. Almost all previous experimental studies have focused
on 1D dielectric multi-layers which provide strong exciton-photon
coupling at the expense of concomitant strong coupling to
{low-quality-factor} leaky modes that render picosecond time-scale polariton lifetimes. 3D PBG materials
enable very strong coupling without the occurrence of leaky
modes. However, structures made of CdTe\cite{cdte} require a very
large number of embedded quantum wells in the active region to provide
sufficiently strong coupling to approach room temperature BEC. Here,
we propose a simple route to above-room-temperature BEC using
TiO$_2$-based photonic crystal architectures. As a reference, we
provide comparison with a corresponding TiO$_2$-SiO$_2$ 1D multi-layer
photonic cavity.

\begin{figure}[]
  \centerline{\includegraphics[height=5.cm]{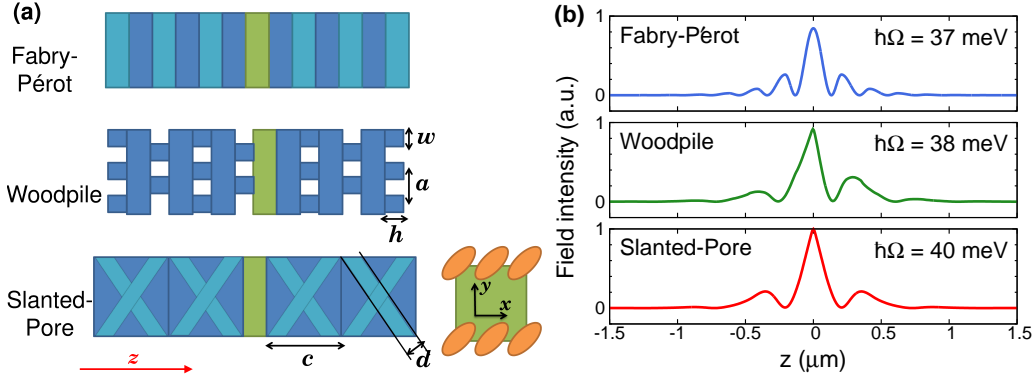}}
  \caption{ {\bf Strong coupling in microcavities} (a) Schematics of the FP, woodpile
    PC and SP PC microcavities. Green regions in the middle represent
    the active layer containing 3 MoSe$_2$ monolayers 
    embedded
    in TiO$_2$ for the PC cavities and SiO$_2$ for the
    FP cavity. For the FP cavity, the dark-blue regions are TiO$_2$ while
    the light-blue regions are SiO$_2$. For woodpile PC cavity the
    white regions denote the air regions while the blue regions are
    TiO$_2$. For SP PC cavity the light-blue regions denote slanted
    air pores while the dark-blue regions are TiO$_2$. Top view of the
    active layer for the SP structure depicts breaking of $x$-$y$
    symmetry due to adjacent air pores, leading to the lifting of
    polariton ground state degeneracy. 
    (b) The distributions of the averaged in-plane photonic 
    field intensity, $S^{-1}_{u.c}\int_{u.c.} d{\vec \rho} (|{\cal
      E}^x_{\vec q}|^2 + |{\cal E}^y_{\vec q}|^2)$ at
    $\vec{Q}=(0,\frac{\pi}{a},0)$, along $z$ direction for the
    woodpile PC and SP PC cavities, and that at $\vec{Q}=0$ for the FP
    cavity. 
  }
  \label{cav}
\end{figure}

In Fig.~\ref{cav} we depict the Fabry-P\'erot (FP), woodpile
photonic crystal (PC) and slanted-pore (SP) PC microcavities, as well
as their cavity-mode field intensities along the growth $z$-direction. There are 3
MoSe$_2$ monolayers in the middle of each microcavity separated by
7~nm TiO$_2$ layers (SiO$_2$ for the FP). Inter-layer Coulomb
correlation energy is suppressed due to the large static dielectric
constant ($\sim 100$) of TiO$_2$ at room temperature\cite{high-d}. The
Coulomb interaction energy for two electrons (holes) separated by $\ge$7~nm
with such high dielectric constant is $\lesssim$2~meV, which is much
smaller than other energy scales such as the exciton-photon coupling
$\hbar\Ome=40$~meV. Consequently, we treat excitons in each monolayer
as independent. The structures of 3 microcavities are illustrated in
Fig.~\ref{cav}a. In the woodpile PC, the height
of each rod is $h=0.3a$ and the width is $w=0.25a$ where $a\simeq
380$~nm is the periodicity in the $x$-$y$ plane (see
Fig.~\ref{cav}a). The SP PC cavity is a SP$_2$
structure of the ${\bf S}/[1,1]\oplus[-1,-1]^{(0.5,0)}$
family\cite{ovi} with the periodicity along the $z$ direction of
$c=1.4a$ where $a\simeq 370$~nm is the in-plane lattice constant. The
diameter of each cylindrical air pore is $d=0.69a$. For
both types of PC microcavity, the thickness of the slab is
0.05$a$. The SiO$_2$-TiO$_2$ FP cavity has two Bragg 
mirrors with periodicity of $a\simeq 210$~nm and a half-wave SiO$_2$
slab of thickness $\simeq 270$~nm in the middle, such that the lowest
band edge cavity mode is close to the exciton recombination energy of
1.55~eV.

Sandwiched PC--quantum-well--PC structures\cite{yang1,yang2} are fabricated in a
layer-by-layer process\cite{noda,noda6}. The PCs and the quantum-well
structure are fabricated independently, the quantum-well structure is
then fused onto the lower PC and finally the upper PC is fused onto
the quantum-well structure\cite{noda,noda6}. Corresponding technologies for
TiO$_2$ PC growth have become mature in the last
decade\cite{noda1,epitaxial,ALD,martin,martin2}. In addition, high quality
MoSe$_2$ thin films can be grown by molecular beam epitaxy\cite{shen}
and chemical vapor deposition\cite{cvd}.

\begin{figure}[]
\centerline{\includegraphics[height=10.cm]{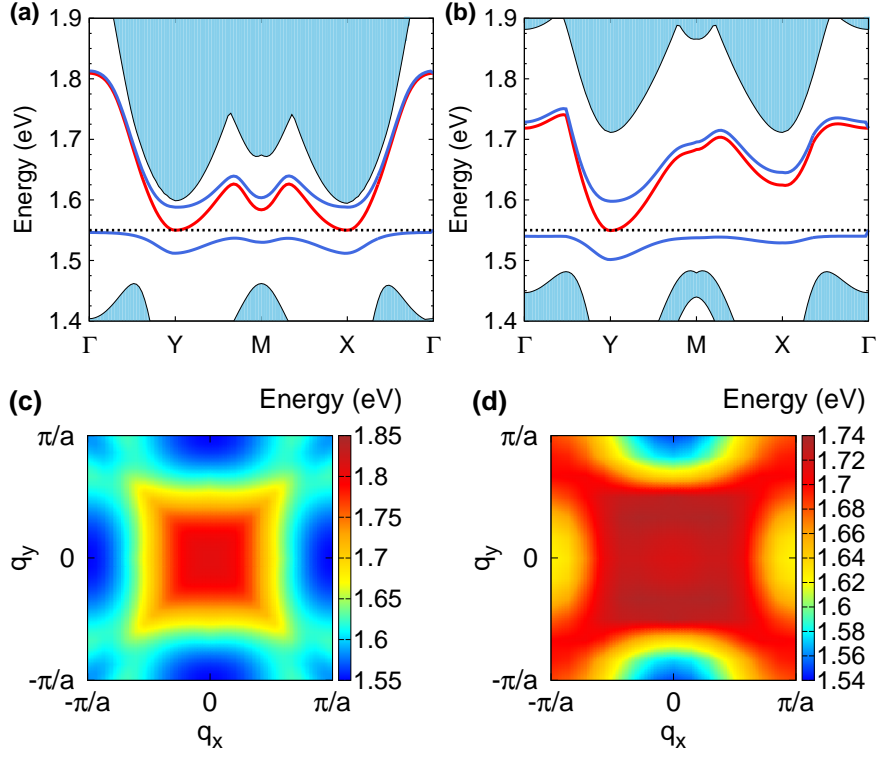}}
  \caption{ {\bf Photonic and polaritonic band structures}. Spectra of photon,
    exciton, and polariton for (a) woodpile and (b) SP PC
    microcavities. Light-blue regions are 3D photonic bands. The red
    curves are the lowest 2D planar guided modes, the dotted curves denote the
    exciton recombination energy, and the blue curves represent the upper and
    lower polariton branches. The lowest band edges of the 2D guided
    (cavity) modes are in resonance with the exciton recombination
    energy, 1.55~eV. This is realized when $a=380$~nm for the woodpile
    and $a=370$~nm for the SP PC. Spectra of the lowest 2D planar
    guided photonic band in 2D wave-vector space for woodpile (c) and SP (d) PC
    microcavities. The polariton ground state is doubly degenerate for
    the woodpile but nondegenerate for the SP microcavity.}
  \label{photon}
\end{figure}

The 3D PBG's in the woodpile and the SP PC's are $\delta
\ome/\ome_c=8.7\%$ and 14\%, respectively 
($\delta\ome$ and $\ome_c$ are the bandwidth and the central frequency
of the PBG, respectively). Photonic band structures near the 3D-PBG
for the woodpile and SP PC cavities reveal several confined 2D guided
bands in the PBG, among which only the lowest one (depicted in
Fig.~\ref{photon}) is relevant for polariton BEC\cite{cdte}.

\section*{Strong exciton-photon coupling}

\begin{figure}[]
\centerline{\includegraphics[height=10cm]{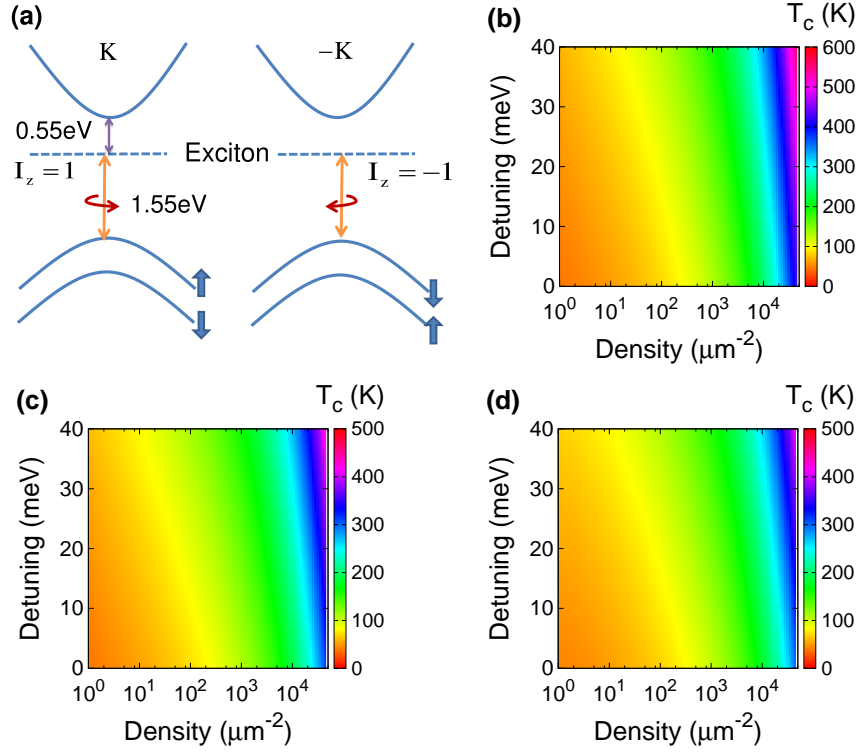}}
  \caption{ {\bf Optical selection rules and polariton BEC transition
      temperatures in MoSe$_2$ cavities} (a) Electronic band structure and
    optical selection rules for MoSe$_2$. The spin-degeneracy in the
    valence band is lifted by spin-orbit interaction. For an excitonic
    optical transition, the $\sigma_+$ photon (angular momentum
    parallel to its wave-vector) is coupled to an exciton in
    the ${\bf K}$ valley consisting of a spin-up $J_z=\frac{3}{2}\hbar$
    hole 
    and a spin-down $J_z=-\frac{1}{2}\hbar$
    electron. 
    The $\sigma_-$ photon is coupled to
    an exciton in the $-{\bf K}$ valley consisting of a spin-down
    $J_z=-\frac{3}{2}\hbar$ hole and a spin-up $J_z=\frac{1}{2}\hbar$
    electron. Polariton BEC transition temperature $T_c$ vs. detuning
    $\Delta$ and polariton density for (b) SP PC, (c) woodpile PC and
    (d) FP microcavities. The side length of the square-shaped
    exciton-trap is $D=5$~$\mu$m.} 
  \label{tc-1}
\end{figure}

The Hamiltonian of the coupled exciton-photon system can be written
as\cite{cdte}
\begin{subequations}
\begin{align}
& H = \sum_{\vec q}\big[ \sum_{l,I_z} E_{X}({\vec q})
  b^\dagger_{l,I_z,\vec q} b_{l,I_z, {\vec
      q}} + \hbar\ome_{\vec q} a_{\vec q}^\dagger
  a_{ \vec q} \big] + H_I , \\
& H_I = \sum_{l,I_z,\vec q}  i\hbar \ov{\Ome}_{l,I_z,\vec q} (b^\dagger_{l,I_z,\vec q} a_{\vec
  q}-a^\dagger_{\vec q}b_{l,I_z,\vec q}) , 
\end{align}
\end{subequations}
where $b^\dagger_{l,I_z,\vec q}$ creates an exciton in the $l$-th
monolayer with total angular momentum $\hbar I_z=\pm \hbar$ along $z$
direction. The energy of the $1s$-exciton is $E_{X}({\vec
  q})=E_{X0}+\frac{\hbar^2q^2}{2m_X}$ where $E_{X0}=1.55$~eV and
$m_{X}$ are exciton emission energy and effective mass, respectively.
In this optical transition from valence to conduction band,
the electron spin is preserved but its orbital angular momentum
changes by $\hbar$. $a_{\vec q}^\dagger$ creates a photon with
wave-vector ${\vec q}$ and frequency $\hbar\ome_{\vec q}$ in the
lowest guided 2D photonic band. The exciton-photon coupling is
\be
\hbar \ov{\Ome}_{l,I_z,\vec q} =  \frac{ d_{cv} |\phi(0)|\sqrt{\hbar \ome_{\vec q}}
}{\sqrt{2 \vep_0}} \sqrt{ \int_{u.c.}
  \frac{d{\vec \rho}}{S_{u.c.}} |u_{I_z,{\vec q}}({\vec \rho},z_l)|^2} , 
\ee
where $d_{cv}=3.6\times 10^{-29}$~Cm (Supplementary Information) is
the absolute value of the interband dipole matrix 
element, $\phi(0)$ is the exciton wavefunction of the $1s$-excitonic
state at zero electron-hole distance, $\ep_0$ is
the vacuum permittivity, $S_{u.c.}=a^2$ is the area of the
unit cell of PC in the $x$-$y$ plane, $z_l$ is the $z$ coordinate of
the $l$-th MoSe$_2$ monolayer plane. $u_{I_z,{\vec q}}=
{\vec {\bf e}}_{I_z}\cdot {\vec u}_{\vec q}$ where ${\vec 
  {\bf e}}_{I_z}=\frac{1}{\sqrt{2}}(1,-i I_z,0)$ is the polarization
vector of the exciton state with $I_z=\pm 1$ and ${\vec u}_{{\vec
    q}}({\vec r})$ is the periodic part of the photonic electric field
${\vec {\cal E}}_{{\vec q}}({\vec
  r}) = \sqrt{\frac{\hbar \ome_{{\vec q}}}{2\vep_0 S}} {\vec u}_{{\vec
    q}}({\vec r}) e^{i {\vec  q}\cdot{\vec \rho}}$. Here $S$ is the
area of the system, ${\vec r}=({\vec \rho},z)$ with ${\vec
  \rho}=(x,y)$ and $S_{u.c.}^{-1}\int_{u.c.}d{\vec \rho}dz 
\ep(\vec{r})|{\vec u}_{{\vec q}}({\vec r})|^2 = 1$.
The special electronic band structure (see Fig.~\ref{tc-1}a) leads to
optical selection rules for the lowest excitonic transitions:
the $\sigma_+$ photon excites only spin-up $J_z=\frac{3}{2}\hbar$ hole
and spin-down $J_z=-\frac{1}{2}\hbar$ electron  in the
${\bf K}$ valley, while $\sigma_-$ photon excites only spin-down
$J_z=-\frac{3}{2}\hbar$ hole and spin-up $J_z=\frac{1}{2}\hbar$ electron in
the $-{\bf K}$ valley\cite{heli1,heli2,yao}. Here $J_z$ is the total
angular momentum along $z$ direction.

The resulting dispersion of the lower polariton branch is
\begin{align}
& E_{lp}({\vec q}) = \frac{E_{X}({\vec q})+\hbar\ome_{\vec q}}{2}
- \left[\left(\frac{E_{X}({\vec q})-\hbar\ome_{\vec
        q}}{2}\right)^2+\hbar^2\Ome_{\vec q}^2\right]^{1/2} .
\end{align}
Here the collective exciton-photon coupling\cite{cdte} is given by
$\Ome_{\vec q}^2 =  \sum_{l,I_z} |\ov{\Ome}_{l,I_z, {\vec q}}|^2$.
Strong light trapping leads to exciton-photon
interaction as large as 40~meV and 38~meV in the SP PC and woodpile PC
microcavities, respectively.

\section*{Above Room Temperature Polariton BEC}

The polariton BEC transition temperature, $T_c$, is calculated by the
criterion\cite{onsager} $N_0/N_{tot}=10\%$ where $N_0$ and $N_{tot}$
are the population on the ground state and the total polariton number
respectively\cite{onsager}. We consider a quantum box trap for
polaritons with box area defined by the finite area of embedded
MoSe$_2$ thin films. As shown below, the trap size, characterized by
the side length $D$ of the square box, does not alter the transition
temperature considerably over a range from a few microns to one
centimeter. The quantum box confinement leads to quantization of
wave-vector and energy. At thermal equilibrium, 
$N_0 = [e^{(E_0-\mu)/(k_BT)}-1]^{-1}$, $N_{tot} = \sum_j
[e^{(E_j-\mu)/(k_BT)}-1]^{-1}$, where $E_0$ is the ground state
energy, $j$ runs over all discrete 
energy levels, and $\mu$ is the chemical potential. As shown
previously\cite{cdte}, $T_c$ is essentially defined by the polariton
dispersion depth, $V_{lp} \equiv \Delta/2  +
\sqrt{\left(\Delta/2\right)^2+\hbar^2\Ome^2}$,
where $\Delta = E_{X0}-\hbar\ome_0$ is the exciton-photon detuning and
$\hbar\Ome\equiv \hbar\Ome_{\vec{Q}^{(\nu)}}$ is the exciton-photon
coupling at photonic band edge. As $k_BT$ approaches $V_{lp}$, the
population on exciton-like states with very large effective mass
becomes significant\cite{cdte}. The pure exciton BEC transition
temperature is calculated to be less than $4$~K for density less than
$(10a_B)^{-2}$ and trap size $D$ of 1 micron in a MoSe$_2$
monolayer. On the other hand for positive detuning $\Delta =30$~meV
and $\hbar\Ome=40$~meV, the dispersion depth is 58~meV. This can
support polariton BEC up to $V_{lp}/k_B\simeq 600$~K.

Figs.~\ref{tc-1}(b)-(d) give the polariton BEC transition temperatures
for different detuning $\Delta$ and polariton densities for the SP PC,
woodpile PC, and FP cavities. We assume that the exciton gas is
initially created by incoherent pumping that (after thermal
relaxation) equally populates all degenerate dispersion minima
($\vec{Q}$-points) in the photonic band structure. As seen in
Fig.~\ref{photon}, the woodpile PC has a doubly degenerate polariton
minimum whereas the polariton ground state is nondegenerate for the
SP PC. The trap size is $D=5~\mu$m. The $T_c$ is enhanced with
increasing detuning as the polariton dispersion depth
increases. The highest $T_c$ is reached at the
largest detuning $\Delta=40$~meV which is 590~K, 450~K, and 430~K for
the SP PC, woodpile PC and FP cavities, respectively. The BEC
transition temperature $T_c$ is highest in the SP PC cavity since it
contains a nondegenerate polariton ground state. This is provided by the
placement of the active layer at a plane that breaks the $x$-$y$
symmetry of the criss-crossing pores above and below (see Fig.~\ref{cav}a). In contrast, there are two
degenerate polariton minima in the woodpile that reduces effective
polariton density for BEC by a factor of two. In the case of the FP
cavity, the polariton density available for BEC is divided by the two
degenerate optical polarization states at ${\vec Q}=0$. We do not
consider larger detuning because a too large $\Delta$ will reduce the 
exciton fraction of polariton as well as the polariton-phonon and
polariton-polartion scattering, which leads to much longer
equilibration time. 

\begin{figure}[]
\centerline{\includegraphics[height=5cm]{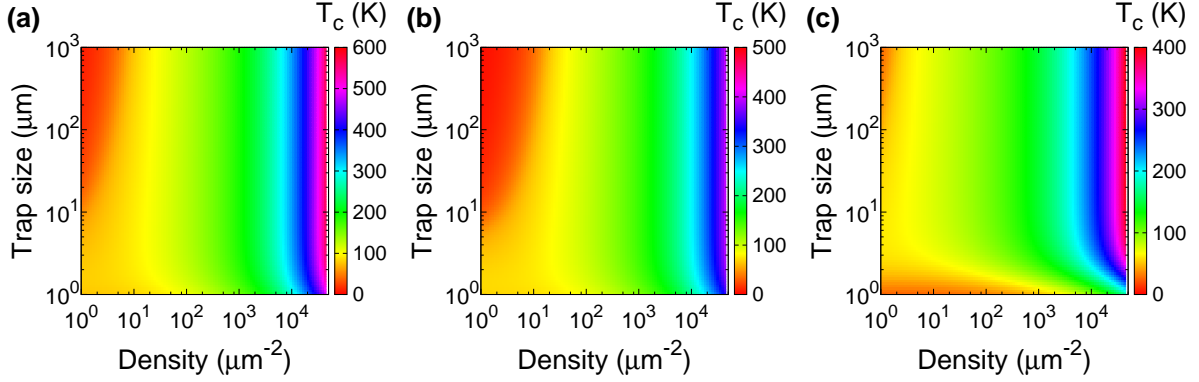}}
  \caption{ {\bf High-temperature polariton BEC} Polariton BEC transition temperature $T_c$
    vs. trap size and polariton density for (a) SP PC, (b) woodpile PC, (c)
    FP microcavities. Detuning is $\Delta=40$~meV,
    corresponding to $\lambda=820$~nm for the FP cavity,
    $a=390$~nm for woodpile PC cavity, and $a=380$~nm for SP PC
    cavity. } 
  \label{tc}
\end{figure}

The polariton BEC transition temperatures at 
$\Delta=30$~meV for different trap sizes $D$ and polariton densities
for each cavity are shown in Fig.~\ref{tc}. The trap size dependence is prominent
for low polariton densities, but very weak for high polariton
densities\cite{cdte}. At low densities, $k_BT_c\ll V_{lp}$ and
polariton equilibrium is governed by the low-lying excited states of
very small effective mass particles in the box trap that are very
sensitive to $D$. At high polariton densities, $k_BT_c\lesssim V_{lp}$ and
a significant fraction of polaritons acquire the bare exciton mass for
which the level spacing is less sensitive to $D$. Eventually $T_c$
tends to zero for very large trap size $D$ (when the low-lying excited states
are extremely close to the ground state) according to the Mermin-Wagner
theorem\cite{wagner}. In all these calculations the largest polariton
density, $5\times 10^4$~$\mu$m$^{-2}$, corresponds to an exciton
density in each MoSe$_2$ monolayer less than $(11a_B)^{-2}$, well below
the exciton saturation density of $(5a_B)^{-2}$\cite{saturation}.

Fabrication induced disorder and exciton line broadening tend to
degrade high-temperature polariton BEC. In a PBG material, light in
the cavity mode is well-protected from scattering into unwanted modes.
Consequently, the role of disorder is primarily to shift the cavity
resonance frequency. For photonic (dielectric)
disorder, our calculation indicates that in the woodpile and SP PC
cavities, photonic structure deviations less than 6~nm in the active
slab layer, less than 12~nm in the logs close to the active slab 
layer, or less than 20~nm in the logs two periods away, do not affect the band edge
of the lowest planar guided mode by more than 2\%. This fluctuation is
equivalent to alteration of the detuning by $\pm$30~meV, which can
reduce the $T_c$ to 360~K (480~K) if the detuning decreases to 10~meV
for the woodpile (SP) PC cavity. Polariton BEC is also found to be
robust to exciton homogeneous and inhomogeneous broadening
although the Rabi-splitting and polariton dispersion depth are
slightly reduced (e.g., Rabi splitting in SP cavity is reduced
from 80~meV to 62~meV) (see Supplementary Information).

In contrast there is no 3D PBG for the FP cavity and light-trapping is
not robust to disorder. Excitons in the FP with wave-vector
$q\gtrsim 0.3 q_0$ with $q_0=E_{X0}/(\hbar c)$ are strongly coupled to
leaky (air) modes, leading to rapid radiative decay\cite{Tassone}.
This decay into leaky modes is not alleviated by increasing the
quality of the FP cavity mode and is exacerbated by imperfect
fabrication accuracy. In real FP cavities, dielectric disorder
scatters polaritons with wave-vector $q\sim 0$ into leaky modes with
the same energy. Such scattering significantly limits the quality
factor of the (nonleaky) cavity mode ($\sim 10^3$). The overall
polariton lifetime due to radiative decay is very short ($\lesssim 1$~ps)
in SiO$_2$/TiO$_2$ multilayers\cite{tio2-fp} and in other\cite{zno2}
FP cavities. {The resulting sub-picosecond polariton lifetime (see
  also calculation in the Supplementary Information) is still comparable with the
  thermalization time, although calculations have shown that
  phonon scattering is very efficient (about 0.1~ps scattering time)
  in MoSe$_2$ at room temperature\cite{mose2-phonon}.} 
In contrast, PC cavities protected by a 3D PBG
enable light-trapping in all directions, leading to polaritons with
lifetime limited only by nonradiative decay {(see Supplementary
  Information). This nonradiative decay is further suppressed by our
  detuning ($\Delta>0$) that renders our polariton more photon-like
  than exciton-like.} The
quantum coherence of polaritons in the cavity can be probed and
manipulated using light propagating through engineered waveguide
channels in the 3D PBG cladding\cite{oc,noda6}.

In the absence of any microcavity, optically excited valley polarization of excitons
can last only about 9~ps in MoSe$_2$ at room temperature\cite{tau-x}. In high-quality PC
microcavities strong light trapping over all directions can enhance
polariton lifetime beyond {130~ps (the exciton lifetime in
MoSe$_2$ monolayer without any cavity at room temperature)\cite{tau-x}.} When
the polaritons form a BEC, optical, valley, and
spin polarization coherences  maintained on such long time-scales
may be useful for quantum information processing
based on manipulation of the spin and valley degrees of freedom\cite{gating,gating2}.

\section*{Conclusions}
We have theoretically demonstrated photonic crystal architectures based
on TiO$_2$ with very strong light-matter interaction and long
polariton lifetime for above-room-temperature, equilibrium
exciton-polariton BEC. The 3D photonic band gap robustly supports
strongly confined guided modes in a slab cavity containing three
separated excitonic MoSe$_2$ monolayers. Unlike 1D periodic
Fabry-P\'erot structures that suffer from substantial radiative decay
through directions other than the stacking direction, the 3D PBG
structures suppress radiative decay in all directions enabling
equilibration and long-time polariton coherence. 
An important feature of the slanted pore PC architecture is that it
provides a non-degenerate lower polariton dispersion minimum, without
recourse to a symmetry-breaking external field\cite{electric}. This
enables a remarkable equilibrium high-temperature Bose-Einstein
condensate, above 500~K, with only three quantum well layers.
Both TiO$_2$ based photonic
crystals and MoSe$_2$ thin films have been fabricated with mature
technologies providing a very accessible route to high-temperature
BEC. This BEC is robust to disorder,
fabrication imperfections, exciton homogeneous and inhomogeneous
line-broadening. The unique coupling among the
optical, valley, and spin degrees of freedom in MoSe$_2$ provides a
condensate with novel internal symmetries that can be manipulated by
external fields. Polariton BEC with long-lived optical, valley, and
spin coherence persisting above room-temperature offers new
opportunities for macroscopic quantum physics.

\section*{Methods}

The photonic band structures and electric field distributions for PC
microcavities are calculated using the plane wave expansion
method, while those for the FP microcavity are calculated using the
transfer matrix method. The refractive indices of TiO$_2$ and SiO$_2$
are 2.7 and 1.5, respectively\cite{material}. In MoSe$_2$ the
effective mass for conduction band electron and valence band hole are,
$m_e=0.7m_0$ and $m_h=0.55m_0$, respectively\cite{x2}. The effective
mass of exciton is $m_X=m_e+m_h$. The exciton Bohr radius $a_B$ and
interband band dipole matrix element $d_{cv}$ are obtained from
existing studies on electronic band structure in MoSe$_2$ and
excitonic optical absorption\cite{shen,yao} (see Supplementary
Information). The calculation of $T_c$ involves summation over
numerous polariton states which is broken into two parts: the
summation over the first 1600 states and the integration over other
higher states. Polariton energy in a square (hard-wall) quantum box is
calculated according to wave-vector quantization.

\section*{Acknowledgments}
We thank J. E. Sipe, J.-L. Cheng, and R. A. Muniz for helpful
discussions. This work was supported by the Natural Sciences and
Engineering Research Council of Canada and the United States
Department of Energy through Contract No. DE-FG02-06ER46347.

\section*{Author Contributions}
J.-H.J. performed calculations and analysis. S.J. guided the
research.


\section*{Additional information}
Competing financial interests: The authors declare no competing
financial interests.

\newpage

\begin{appendix}

\begin{center}
{\bf SUPPLEMENTARY INFORMATION}
\end{center}

\section*{Guided 2D Photonic band Edge}

In the woodpile PC cavity the lowest 2D photonic band has two dispersion 
minima: the $X$ point ${\vec Q}^{(X)}=(\frac{\pi}{a}, 0, 0)$ and the
$Y$ point ${\vec Q}^{(Y)}=(0, \frac{\pi}{a}, 0)$. The
dispersion has a $C_{4v}$ symmetry, inherited from the $D_{2d}$
symmetry of the woodpile PC cavity. We consider the
situation when the 2D photonic band edge is close to the exciton
emission energy (i.e., $a=380$~nm) where the dispersion of the
2D photonic band near the minima $\nu=X,Y$ is $\hbar\ome_{\vec q} =
\hbar\ome_{0} + \frac{\hbar^2(q_x-Q_x^{(\nu)})^2 }{2m_x^{(\nu)}}
+ \frac{\hbar^2(q_y-Q_y^{(\nu)})^2}{2m_y^{(\nu)}}$. Here,
$m_x^{(X)} = m_y^{(Y)}\simeq 2.4\times 10^{-5}m_0$ and $m_x^{(Y)} =
m_y^{(X)}\simeq 7.2\times 10^{-5}m_0$ where $m_0$ is the bare electron
mass in vacuum. In SP PC cavity the dispersion of the lowest 2D
photonic band has the $C_{2v}$ symmetry. There is only one dispersion
minimum at the $Y$ point  where the
effective mass is also anisotropic with $m_x =5.7\times 10^{-5}m_0$
and $m_y =2.1\times 10^{-5}m_0$ for $a=370$~nm. The effective mass of
photon in the FP cavity is $0.68\times 10^{-5}m_0$ at exciton-photon
resonance.

\section*{Electronic band structure of MoSe$_2$}

Schematic of electronic band structure of MoSe$_2$ monolayer
is shown in Fig.~3a in the main text. Band extrema are located at the
${\bf K}=(\frac{4\pi}{3a_s},0)$ ($a_s$ is the honeycomb lattice
constant of MoSe$_2$) and $-{\bf K}$ points\cite{yao}. Very large splitting 
occurs between spin-up and -down states of 180~meV\cite{mose2} in the
valence band  due to spin-orbit interaction. Denoting the
$z$-components of total, orbital and spin angular momenta as $J_z$,
$L_z$ and $S_z$, the valence band maximum
at the ${\bf K}$ valley gives spin-up $J_z=\frac{3}{2}\hbar$ holes
(absence of an electron with $L_z=-\hbar$ and
$S_z=-\frac{1}{2}\hbar$ in the valence band). Its time-reversal 
partner $-{\bf K}$ valley gives spin-down $J_z=-\frac{3}{2}\hbar$
holes. This special electronic band structure leads to 
optical selection rules for the lowest excitonic states (see
Fig.~3a in the main text): the $\sigma_+$ photon excites only spin-up
$J_z=\frac{3}{2}\hbar$ hole 
and spin-down $(J_z,L_z,S_z)=(-\frac{1}{2}\hbar,0,-\frac{1}{2}\hbar)$
electron  in the 
${\bf K}$ valley, while $\sigma_-$ photon excites only spin-down
$J_z=-\frac{3}{2}\hbar$ hole and spin-up
$(J_z,L_z,S_z)=(\frac{1}{2}\hbar,0,\frac{1}{2}\hbar)$ electron in the 
$-{\bf K}$ valley\cite{yao}.

\section*{Estimation of the exciton-photon coupling in MoSe$_2$}

From Ref.~\cite{book} the imaginary part of the optical susceptibility
at the energy of the $1s$-exciton state $E_{X0}=\hbar\ome_{X0}$ is
given by 
\be
\Im [\chi(\ome_{X0})] =  \frac{2d_{cv}^2|\phi(0)|^2}{L_c\hbar\delta} ,\label{a1}
\ee
where $L_c$ is the quantum well width and $\hbar \delta$ is the
homogeneous broadening of the $1s$-exciton state. From measurements of
$\Im[\chi(\ome_{X0})]$ we deduce $d_{cv}|\phi(0)|$.

The absorptance for an isolated two-dimensional system with thickness
$L_c$ is
\begin{eqnarray}
  \alpha = \frac{ \omega L_c \text{Im}[\chi(\omega)]}{ n \vep_0 c} ,\label{a2}
\end{eqnarray}
where $\omega$ and $c$ are the frequency and speed of light,
respectively, and $n$ is the refractive index. From Ref.~\cite{x2} the
in-plane dielectric constant is 4.7 for a monolayer
MoSe$_2$. Accordingly, we choose the refractive index $n\simeq 2.2$.

For MoSe$_2$ the absorptance $\alpha\simeq 6\%$\cite{shen}. It
follows from Eqs.~(\ref{a1}) and (\ref{a2}) that
\be
d_{cv}|\phi(0)| = \sqrt{ \frac{0.06 n \vep_0 c \hbar\delta}{ 2 \ome_{X0}} } .
\ee
From Ref.~\cite{shen} the homogeneous broadening at
room-temperature is about 50~meV. Therefore,
$d_{cv}|\phi(0)|\simeq 2.4 \times 10^{-20}$~C which is 2.4 times as
large as that of excitons GaAs quantum-wells of width 3~nm ($\simeq
1.0\times 10^{-20}$~C). Using $|\phi(0)|=\sqrt{2/(\pi
  a_B^2)}$ for the 2D exciton and the Bohr radius
$a_B=1.2$~nm\cite{shen}, we estimate $d_{cv}=3.6\times
10^{-29}$~Cm. This is also comparable with\cite{yao} 
$d_{cv}=\sqrt{2}a_ste/E_g=3.4\times 10^{-29}$~Cm where
$a_s=3.313$~\AA\cite{yao} is the honeycomb lattice constant of
MoSe$_2$, $t=0.94$~eV\cite{yao} is the effective electron hopping
parameter, $e=1.602\times 10^{-19}$~C is the electronic
charge and $E_g=2.1$~eV\cite{shen} is the electronic band gap of MoSe$_2$.

\section*{Ultra-short polariton radiative lifetime in Fabry-P\'erot microcavities}

\begin{figure}[]
  \centerline{\includegraphics[height=4.8cm]{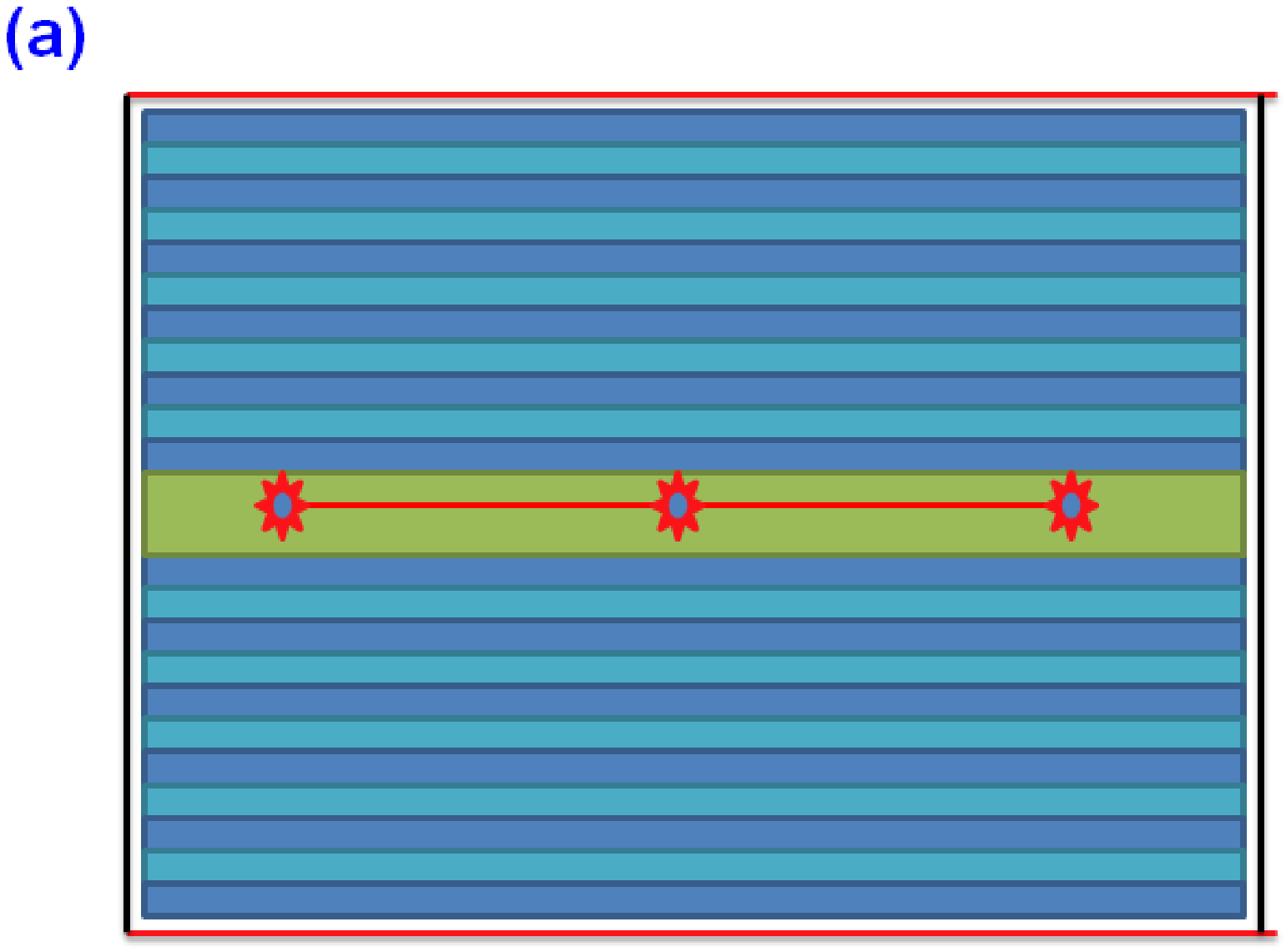}\includegraphics[height=4.8cm]{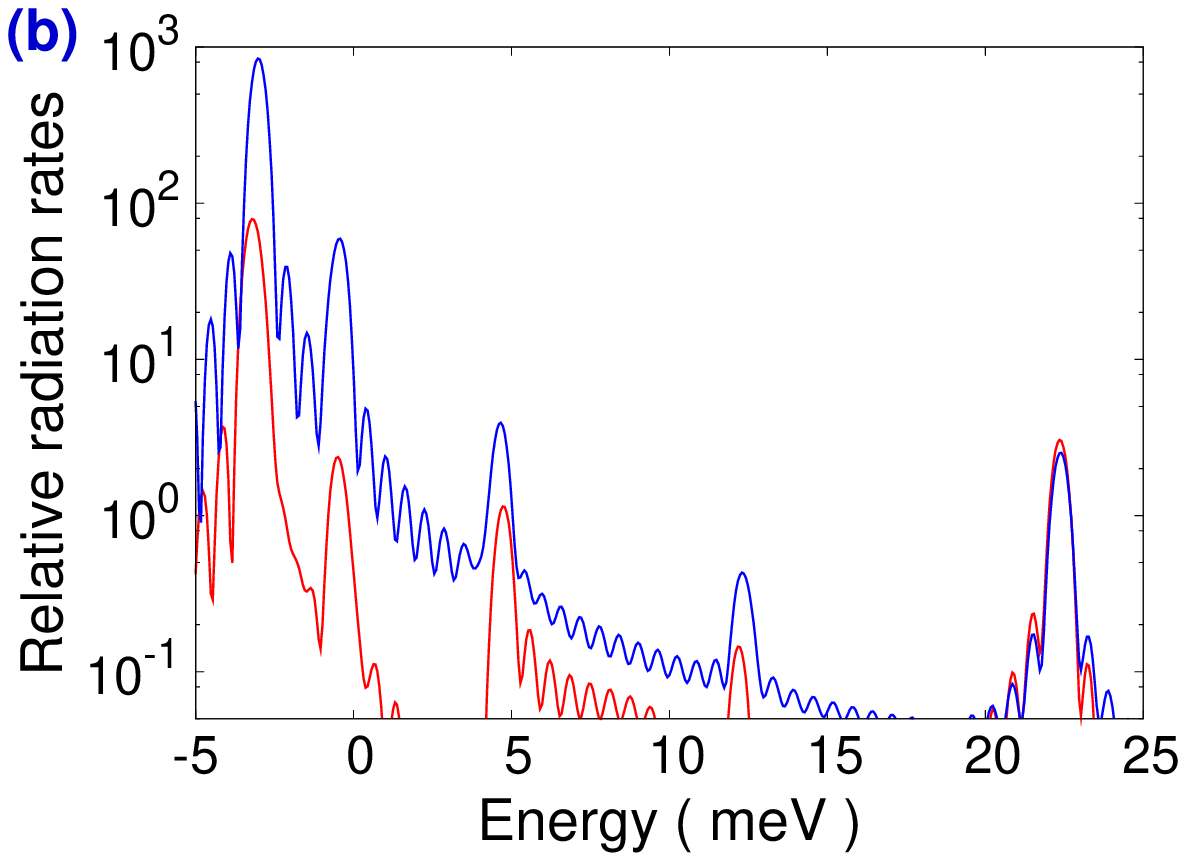}}
  \centerline{\includegraphics[height=4.8cm]{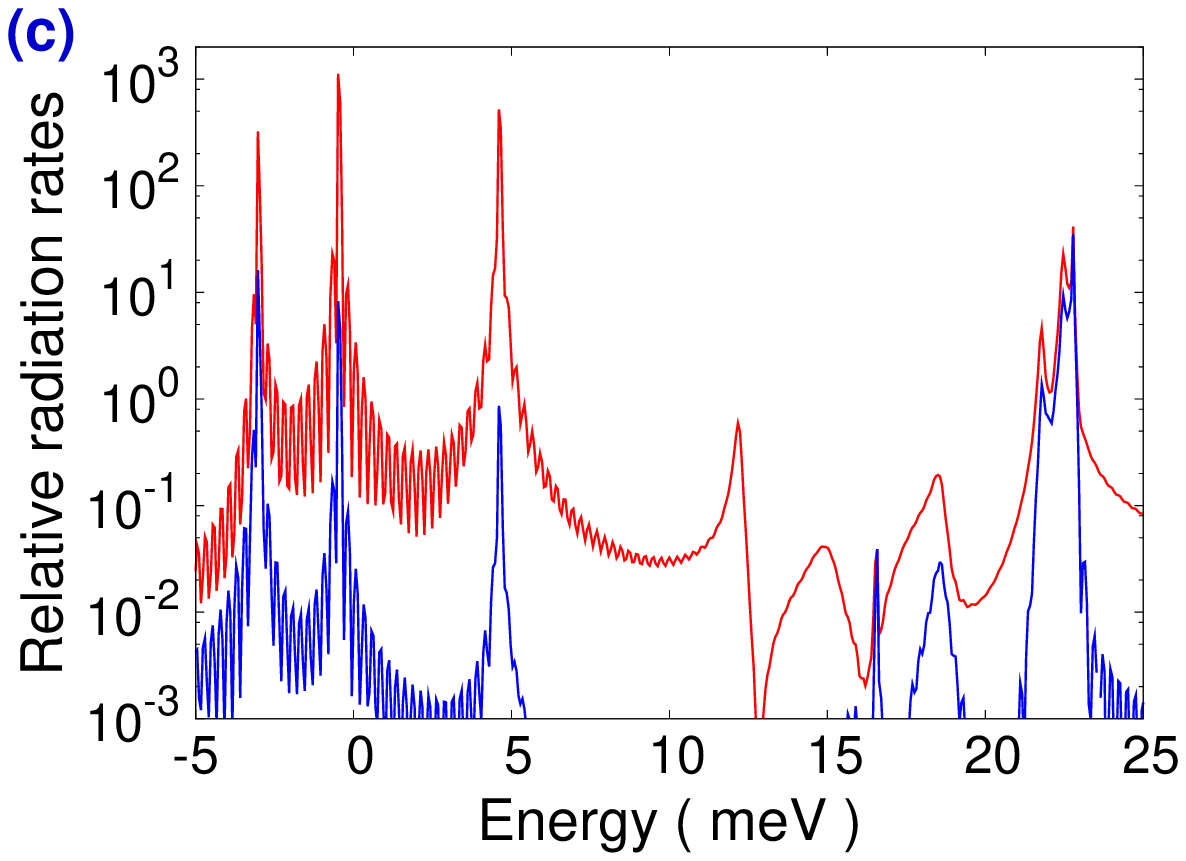}\includegraphics[height=4.8cm]{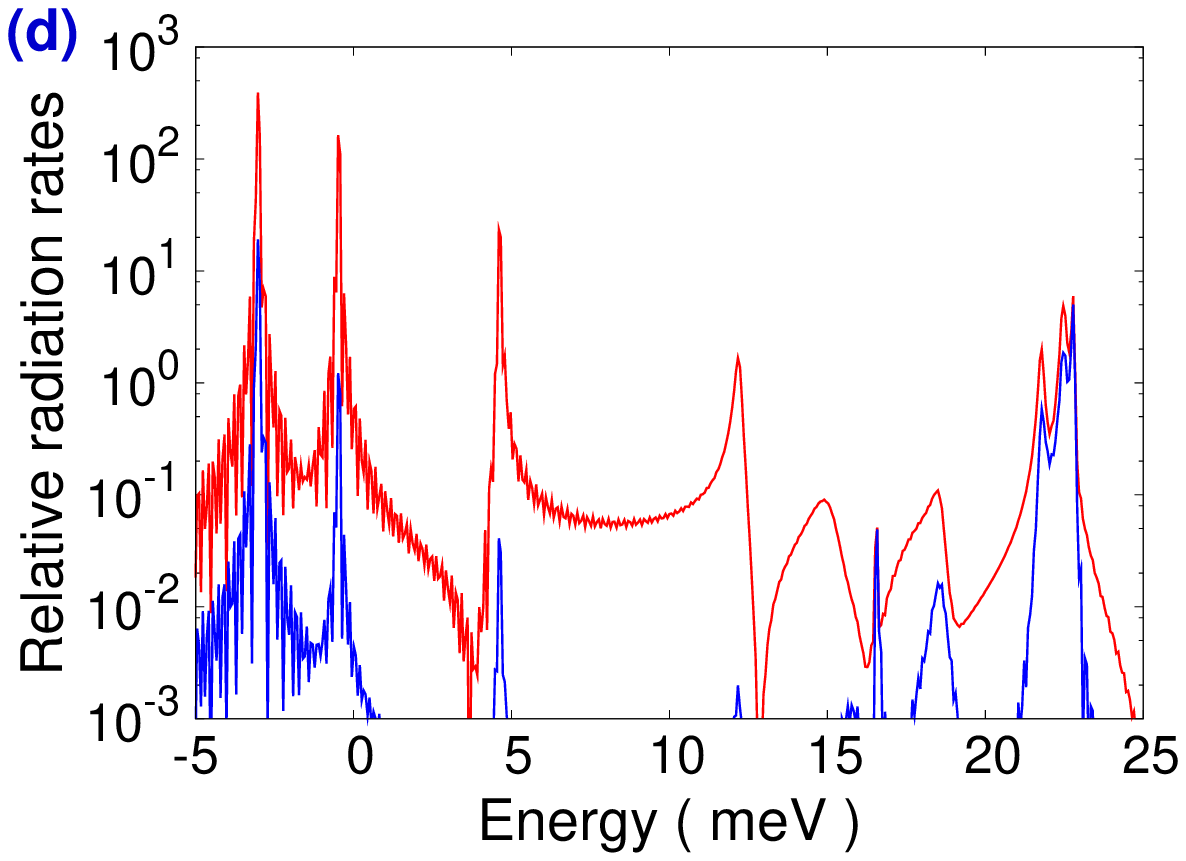}}
  \caption{ Supplementary Figure S1. (a) Structure for 2D FDTD calculation of radiation rate in FP
    cavities. An extended dipole line source with length smaller than the
    width of the FP cavity is placed in the center (denoted by the
  line connecting the three star symbols). A Gaussian pulse excites
  dipole oscillations perpendicular to the paper plane causing outward
  radiation. The radiation flux is
  collected at the boundaries surrounding the cavity [denoted by
  the two red lines (top and bottom) and black lines (left and
  right)]. The radiation flux is calculated for the FP cavity as well
  as in the absence of the Bragg mirrors. The ratio of the
  former over the latter is plotted in (b)-(d). This ratio is resolved
  into escape through the top and bottom of the cavity (blue curves)
  and escape through the side directions (red curves). In the
  simulation, the width of the cavity is $14\lambda$, while the 
  length of the dipole is $12\lambda$ or $4\lambda$. The 
  structure is surrounded by an air region of thickness $\lambda$
  followed by perfectly matching layers of thickness $4\lambda$. $\lambda=0.8$~$\mu$m
  is the length unit and $hc/\lambda$ is equal to exciton emission
  energy $E_{X0}=1.55$~eV. (b)-(d): Dipole radiation rate in FP
  cavities relative to that in vacuum (red curves) as a function of
  the detuning of the dipole (exciton) energy from the Fabry-P\'erot
  cavity mode energy. (b) FP cavity 
  with 8 pairs of $\lambda/4$ SiO$_2$ and TiO$_2$ layers in the
  distributed Bragg mirrors above and below the $\lambda/2$ slab. (c)
  and (d) FP cavities with 15 pairs of $\lambda/4$ layers above and
  below the slab for different dipole sizes. Simulation is done via
  2D FDTD with horizontal dipoles of lengths: (b) and (d) 
  $9.6$~$\mu$m, (c) $3.2$~$\mu$m.}
  \label{fdtd}
\end{figure}

1D periodic microstructures can provide strong coupling to a single
high-quality factor optical mode. However the concomitant strong
coupling to degenerate leaky modes severely reduces the polariton
radiative lifetime. Here, we simulate polariton radiative decay from a finite-size
polariton condensate using an extended dipole at the center of a FP
cavity by finite-difference-time-domain (FDTD). Different dipole sizes
are used to represent small and large polariton BEC trap sizes. The
radiative decay rate in the FP cavity is compared with 
that of the same dipole in vacuum (i.e., without the
cavity). Experiments\cite{tau-x} suggest that the latter is about
1/(130~ps) in a MoSe$_2$ monolayer at room temperature. 2D FDTD
simulations are used to provide a semi-quantitative prognosticator of
a realistic 3D structure (see Fig.~\ref{fdtd}). The radiative decay rate (relative
to that in vacuum) for SiO$_2$-TiO$_2$ FP cavities with different
dipole sizes and numbers of $\lambda/4$ layers are plotted as a
function of the detuning of the dipole (exciton) energy from the
Fabry-P\'erot cavity energy in Fig.~\ref{fdtd}. The radiative decay is
greatly enhanced at some energies while suppressed at other energies
all within the exciton homogeneous line-width. The former
are the hot-spots for exciton radiative decay leading to very short
($\lesssim 1$~ps) polariton lifetime. In Fig.~\ref{fdtd} the zero of
energy $hc/\lambda=E_{X0}=1.55$~eV and all energy scales in the
figures are within the 1D photonic stop-gap at normal incidence.

For FP cavities with 8 pairs of $\lambda/4$ SiO$_2$ and TiO$_2$ layers
(as in Fig.~\ref{fdtd}a) the cavity mode has a quality factor of 
$1.2\times 10^4$, implying a cavity photon lifetime of 11~ps. However,
excitons can decay radiatively into a broad range of very low quality
leaky modes. This radiative decay is most rapid into leaky modes with
energy lower than the cavity mode. The $10^3$ enhancement of radiation
rate leads to sub-picosecond polariton lifetime. With 15 pairs of
$\lambda/4$ layers the FP cavity mode has a quality factor as high as
$1.8\times 10^6$ implying a cavity photon lifetime of 1500~ps. As
before, excitons can radiate through leaky modes (Figs.~\ref{fdtd}b
and \ref{fdtd}c) emerging from the sides of the FP. Radiation through
leaky modes becomes weaker for the larger dipole
size of $12\lambda$ than for the smaller dipole size of $4\lambda$. Nevertheless,
radiation lifetimes at the hot-spots are on the order of 1~ps. In the
limit of an infinitely extended dipole contained in an FP cavity of
infinite lateral extent, momentum conservation requires that all
radiation is emitted into the cavity mode. However, in this case, BEC
is excluded by the Mermin-Wagner theorem\cite{wagner}.

Dielectric disorder in realistic SiO$_2$-TiO$_2$ FP's reduces the
polariton lifetime even below our calculated time
scales\cite{tio2-fp}. Moreover, at room temperature, excitons
with energies away from strongly-coupled leaky modes but within the
homogeneous linewidth (50~meV\cite{shen}) and phonon energy
(43~meV\cite{LO-mose1}) can be scattered efficiently (within $\lesssim
1$~ps) by phonons\cite{LO-mose2} to the leaky modes and then
radiate rapidly. These effects reduce polariton lifetime to 
$\lesssim 1$~ps even for FP cavities with 15 pairs of $\lambda/4$ layers
and cavity quality factor of $1.8\times 10^6$. {The sub-picosecond
  lifetime of polaritons is still on the same order of magnitude with the phonon
  scattering time (about 0.1~ps) in MoSe$_2$ at room
  temperature\cite{LO-mose2}.} In contrast, in a 3D
PBG microcavity architecture with a microcavity embedded in a 3D PBG
material\cite{noda,noda6}, all leaky modes can be eliminated leading
to very long polariton lifetime, limited only by exciton nonradiative
decay. These nonradiative decay channels include Auger recombination
arising from exciton-exciton collision and
Shockley-Read-Hall recombination arising from electronic defects in
the active semiconductor monolayers. Auger recombination is usually
very inefficient in moderate to large band gap semiconductors such as
MoSe$_2$ for the moderate exciton densities we consider 
(volume density less than $10^{18}$~cm$^{-3}$)\cite{zhao,cdte}. Recently high
(electronic) quality, large area MoSe$_2$ monolayers were
successfully prepared by various methods\cite{cvd,method2,method3},
{exploiting the significant mechanical, chemical and thermal
  stability of MoSe$_2$ monolayers\cite{method3}. In our architecture,
  the MoSe$_2$ monolayers are encapsulated by TiO$_2$ layers above and
  below, protecting them from defects introduced by wafer-fusion of
  the central slab to the photonic crystal.}
In well-fabricated samples, the polariton lifetime in our 3D PBG
microcavities can be much longer than 130~ps, while efficient phonon
scattering facilitates equilibration of the polariton gas in $\lesssim
1$~ps\cite{LO-mose2}.

\section*{Effect of exciton inhomogeneous and homogeneous broadening}

\begin{figure}[]
  \centerline{\includegraphics[height=4.8cm]{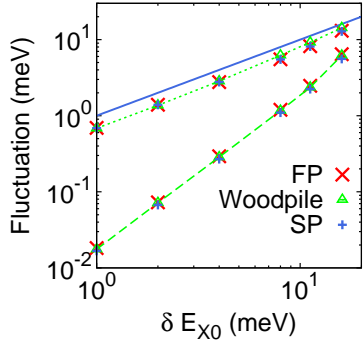}}
  \caption{Supplementary Figure S2.  Root mean square deviation of the vacuum Rabi splitting (points linked
    by dotted curve) and the dispersion depth of polariton $V_{lp}$
    (points linked by dashed curve) as functions of the root mean
    square deviation of the exciton emission energy $\delta E_{X0}$
    when the detuning is
    $\Delta=0$. The solid reference line denotes fluctuation equal to the
    exciton inhomogeneous broadening $\delta E_{X0}$.} 
  \label{inhom}
\end{figure}

As a result of polariton motional narrowing,
inhomogeneous broadening within a given monolayer has only a minor effect on
BEC\cite{yang1,yang2,book-kavo}. Inhomogeneous
broadening between different monolayers is modeled by a 
Gaussian random shift of the exciton emission
energy in each monolayer, $E_{X0}\to E_{X0}+\delta E_l$, $l=1,2,3$.
The root mean square deviation of the vacuum Rabi
splitting and the polariton dispersion depth $V_{lp}$,
calculated as a function of exciton inhomogeneous broadening $\delta
E_{X0}$, are shown in Fig.~\ref{inhom}. The results for the three types of
cavity are almost the same. The fluctuation of the vacuum Rabi
splitting is considerably smaller than the exciton inhomogeneous
broadening due to averaging among different
monolayers\cite{cdte}, whereas the fluctuation
of the dispersion depth $V_{pl}$ is comparable to $\delta
E_{X0}$. Nevertheless, for inhomogeneous broadening less than 10~meV,
the dispersion depth remains larger than 48~meV enabling BEC up to
500~K (370~K) for the SP (woodpile) PC cavity.

Exciton homogeneous broadening is considerable in MoSe$_2$ at room
temperature due to phonon
scattering\cite{LO-mose2}. Optical measurements reveal
a broadening of 50~meV in
Ref.~\cite{shen}. Picosecond phonon scattering\cite{LO-mose2} facilitates thermal
equilibration of polaritons within  several
picoseconds. Homogeneous broadening slightly degrades the vacuum Rabi
splitting and the polariton dispersion depth. If the 
homogeneous broadening is described by a imaginary part of exciton energy,
$i\Gamma$, the vacuum Rabi splitting is reduced
from $2\hbar\Ome=80$~meV to $2\sqrt{\hbar^2\Ome^2-\Gamma^2/4}=62$~meV
at zero detuning for $\Gamma=50$~meV. For a detuning of $\Delta=40$~meV, exciton
homogeneous broadening of $\Gamma = 50$~meV reduces the
lower-polariton dispersion depth $V_{lp}$ by 5.5~meV. The highest transition
temperature is then reduced from 590~K to 560~K. If an inhomogeneous
broadening of 30~meV is taken into account as well, the transition
temperature will be reduced to 430~K which is still above
room-temperature. This suggests that
polaritons in woodpile and SP PC microcavities are robust against
electronic and photonic disorder and exciton homogeneous broadening,
enabling above room-temperature polariton BEC in realistic systems.

\end{appendix}


\begin{thebibliography}{99}

\bibitem{mc} Yamamoto, Y., Tassone, T. \& Cao, H. {\it Semiconductor
    cavity quantum electrodynamics}. (Springer, Berlin, 2000).

\bibitem{bec1} Deng, H., Weihs, G., Santori, C., Bloch, J.,
  \& Yamamoto, Y. Condensation of semiconductor
microcavity exciton polaritons. {\sl Science} {\bf 298}, 199-202 (2002).

\bibitem{bec2} Kasprzak, J. {\sl et al.} 
 Bose-Einstein condensation of exciton
 polaritons. {\sl Nature} {\bf 443}, 409-414 (2006). 

\bibitem{bec3} Balili, R., Hartwell, V., Snoke, D., Pfeiffer, L.,
  West, K. Bose-Einstein condensation of microcavity polaritons in a
  trap. {\sl Science} {\bf 316}, 1007-1010 (2007).

\bibitem{bec4} Kasprzak, J., Solnyshkov, D. D., Andr\'e, R., Dang,
  L. S. \& Malpuech, G. Formation of an exciton polariton condensate:
  thermodynamic versus kinetic regimes. {\sl Phys. Rev. Lett.} {\bf
    101}, 146404 (2008).



\bibitem{deng} Deng, H., Haug, H. \& Yamamoto, Y. Exciton-polariton
  Bose-Einstein condensation. {\sl Rev. Mod. Phys.} {\bf 82}, 1489-1537 (2010).






\bibitem{zno} Lai, Y.-Y., Lan, Y.-P. \& Lu, T.-C. Strong light-matter interaction in ZnO
  microcavities. {\sl Light Sci. Appl.} {\bf 2}, e76 (2013).

\bibitem{zno2}
  Li, F. {\sl et al.}  From excitonic to photonic polariton
    condensate in a ZnO-based microcavity. {\sl Phys. Rev. Lett.} {\bf 110}, 196406 (2013).

\bibitem{gan} Christopoulos, S. {\sl et al.} 
 Room-temperature polariton lasing in
 semiconductor microcavities.  {\sl Phys. Rev. Lett.} {\bf 98}, 126405
(2007). 

\bibitem{gan2}
   Christmann, G., Butt\'e, R., Feltin, E.,
  Carlin, J.-F. \& Grandjean, N. Room temperature polariton
  lasing in a GaN/AlGaN multiple quantum well microcavity.
  {\sl Appl. Phys. Lett.} {\bf 93}, 051102 (2008). 

\bibitem{gan3}  Levrat, J. {\sl et al.} Condensation phase diagram of
  cavity polaritons in GaN-based microcavities: experiment and
  theory. {\sl Phys. Rev. B} {\bf 81}, 125305 (2010).

\bibitem{polymer} Plumhof, J. D., St\"oferle, T., Mai, L., Scherf, U. \&
  Mahrt, R. F. Room-temperature Bose-Einstein condensation of
    cavity exciton-polaritons in a polymer. {\sl Nat. Mater.} {\bf 13}, 247-252 (2014).


\bibitem{rt} Snoke, D.
  Microcavity polaritons: a new type of light switch. {\sl Nat.
  Nanotechnol.} {\bf 8}, 393-395 (2013).



\bibitem{pbg} John, S. Strong localization of photons in certain
  disordered dielectric superlattices. {\sl Phys. Rev. Lett.} {\bf 58},
  2486-2489 (1987).

\bibitem{pbg2}
  Yablonovitch, E.  Inhibited spontaneous emission in
  solid-state physics and electronics. {\sl Phys. Rev. Lett.} {\bf 58}, 2059-2062 (1987).





\bibitem{shen} Ugeda, M. M. {\sl et al.} Observation of giant
  bandgap renormalization and excitonic effects in a monolayer
  transition metal dichalcogenide semiconductor. arXiv:1404.2331, {\sl Nat. Mater.} in press. 


\bibitem{2dphc1} Gan, X. {\sl et al.} Controlling the spontaneous
  emission rate of monolayer MoS$_2$ in a photonic crystal nanocavity. 
    {\sl Appl. Phys. Lett.} {\bf 103}, 181119 (2013).

\bibitem{2dphc2} Wu, S. {\sl et al.}  Control of two-dimensional
  excitonic light emission via photonic crystal. {\sl 2D Mater.} {\bf 1},
  011001 (2014).



\bibitem{mos2-fp} Liu, X. {\sl et al.} 
Strong   light-matter coupling in two-dimensional atomic
crystals. {\sl arXiv}:1406.4826 


\bibitem{mose2} Ross, J. S. {\sl et al.} Electrical control of
  neutral and charged excitons in a monolayer semiconductor. {\sl Nat. Commun.} {\bf 4}, 1474 (2013).

\bibitem{cdte}  Jiang, J. H. \& John, S. Photonic crystal architecture for room temperature equilibrium
  Bose-Einstein condensation of exciton-polaritons. {\sl Phys. Rev. X} {\bf
  4}, 031025 (2014).



\bibitem{yao} Xiao, D., Liu, G.-B., Feng, W., Xu, X. \& Yao, W. 
  Coupled spin and valley physics in monolayers of MoS$_2$ and other
  group-VI dichalcogenides.
  {\sl Phys. Rev. Lett.} {\bf 108}, 196802 (2012).





\bibitem{electric} Schneider, C. {\sl et al.} 
An electrically pumped polariton
  laser. {\sl Nature} {\bf 497}, 348-352 (2013)




\bibitem{gating} Amo, A. {\sl et al.} 
   Exciton-polariton
   spin switches. {\sl Nature Photon.} {\bf 4}, 361-366  (2010)

\bibitem{gating2} Cerna, R. {\sl et al.}
 Ultrafast tristable spin
 memory of a coherent polariton gas. {\sl Nat. Commun.} {\bf
    4}, 2008 (2013).




\bibitem{high-d} Parker, R. A. Static dielectric constant of
    rutile (TiO$_2$), 1.6-1060~K. {\sl Phys. Rev.} {\bf 124}, 1719-1722 (1961).

\bibitem{ovi} Toader, O., Berciu, M., \& John, S.  Photonic band
    gaps based on tetragonal lattices of slanted pores. {\sl Phys. Rev. Lett.} {\bf
    90}, 233901 (2003).


\bibitem{yang1} John, S. \& Yang, S.  Electromagnetically induced
  exciton mobility in a photonic band gap. {\sl Phys. Rev. Lett.} {\bf 99},
  046801 (2007).

\bibitem{yang2}  Yang, S. \& John, S. Exciton dressing and capture
  by a photonic band edge. {\sl Phys. Rev. B} {\bf 75}, 235332 (2007).


\bibitem{noda} Ogawa, S., Imada, M., Yoshimoto, S., Okano, M., \&
  Noda, S. Control of light emission by 3D photonic
    crystals. {\sl Science} {\bf 305}, 227-229 (2004)

\bibitem{noda6}
 Noda, S., Fujita, M., \& Asano, T. Spontaneous-emission control by photonic crystals
 and nanocavities. {\sl Nature Photon.} {\bf 1}, 449-458 (2007).


\bibitem{noda1} Noda, S., Tomoda, K., Yamamoto, N. \& Chutinan, A. 
  Full three-dimensional photonic bandgap crystals at near-infrared
  wavelengths. {\sl Science} {\bf 289}, 604-606 (2000). 


\bibitem{epitaxial} Nelson, E. {\sl et al.}  Epitaxial growth
    of three-dimensionally architecture optoelectronic devices. {\sl Nat. Mater.} {\bf
    10}, 676-681 (2011).

\bibitem{ALD}  King, J. S. {\sl et al.}  Infiltration and inversion
  of holographically defined polymer photonic crystal templates by
  atomic layer deposition. {\sl Adv. Mater.} {\bf 18}, 1561-1565
  (2006).

\bibitem{martin} Deubel, M., Wegener, M., Kaso, A. \& John, S. 
    Direct laser writing and characterization of ``slanted-pore''
    photonic crystals. {\sl Appl. Phys. Lett.} {\bf 85}, 1895-1897
    (2004).

\bibitem{martin2}
    Deubel, M. {\sl et al.}  Direct laser writing of
    three-dimensional photonic-crystal templates for
    telecommunications. {\sl Nat. Mater.} {\bf 3}, 444-447 (2004).

\bibitem{cvd} Shim, G. W. {\sl et al.}  Large-area single-layer
  MoSe$_2$ and its van der Waals heterostructures. {\sl ACS Nano} {\bf 8},
  6655-6662 (2014).








\bibitem{heli1} Zeng, H.,  Dai, J., Yao, W., Xiao, D. \& Cui, X. Valley polarization in MoS$_2$ monolayers by
optical pumping. {\sl Nat. Nanotechnol.} {\bf 7}, 490-493 (2012).


\bibitem{heli2} Mak, K. F., He, K., Shan, J. \& Heinz, T. F. 
Control of valley polarization in monolayer MoS$_2$ by optical
helicity. {\sl Nat. Nanotechnol.} {\bf 7}, 494-498 (2012).



\bibitem{onsager}  Penrose, O. \& Onsager, L.  Bose-Einstein
  condensation and liquid helium. {\sl Phys. Rev.} {\bf 104}, 576-584
  (1956). 

\bibitem{wagner} Mermin, N. D. \&  Wagner, H. Absence of
  Ferromagnetism or antiferromagnetism in one- or two-Dimensional
  isotropic Heisenberg models. {\sl Phys. Rev. Lett.} {\bf
    17}, 1133-1136 (1966).


\bibitem{saturation} Schmitt-Rink, S.,  Chemla, D. S. \&
   Miller, D. A. B.  Theory of transient excitonic optical
   nonlinearities in semiconductor quantum-well structures. {\sl Phys. Rev. B} {\bf 32}, 6601-6609 (1985).





\bibitem{Tassone}  Savona, V. \& Tassone, F. Exact Quantum
  Calculation of polariton dispersion in semiconductor
  microcavities. {\sl Solid State Commun.} {\bf 95}, 673-678 
  (1995).

\bibitem{tio2-fp} Bhattacharya, P. {\sl et al.} 
 Room temperature
    electrically injected polariton laser. {\sl Phys. Rev. Lett.} {\bf 112},
  236802 (2014).


\bibitem{mose2-phonon} Jin, Z., Li, X., Mullen, J. T. \& Kim,
  K. W. Intrinsic transport properties of electrons and holes in
  monolayer transition-metal dichalcogenides. {\sl Phys. Rev. B} {\bf
    90}, 045422 (2014).




\bibitem{oc} 
  Chutinan, A., John, S. \& Toader, O. Diffractionless
  flow of light in all-optical microchips. {\sl Phys. Rev. Lett.} {\bf
    90}, 123901 (2003).








\bibitem{tau-x} Kumar, N., He, J., He, D., Wang, Y. \& Zhao, H. Valley
  and spin dynamics in MoSe$_2$ two-dimensional crystals. {\sl
    Nanoscale}, {\bf 6}, 12690 (2014).


\bibitem{material}  Landmann, M., Rauls, E. \& Schmidt, W. G.  The
  Electronic structure and optical response of rutile, anatase and
  brookite TiO$_2$. {\sl J. Phys.:
  Condens. Matter} {\bf 24}, 195503 (2012).


\bibitem{x2} Ramasubramaniam, A. Large excitonic effects in
  monolayers of molybdenum and tungsten dichalcogenides. {\sl Phys. Rev. B} {\bf 86}, 115409 (2012).














\end{thebibliography}

\begin{thebibliography}{99}



\bibitem{yao} Xiao D., Liu, G.-B., Feng, W., Xu, X. \& Yao, W. 
  Coupled spin and valley physics in monolayers of MoS$_2$ and other
  group-VI dichalcogenides.
    {\sl Phys. Rev. Lett.} {\bf 108}, 196802 (2012).



\bibitem{mose2} Ross, J. S. {\sl et al.} Electrical control of
  neutral and charged excitons in a monolayer semiconductor. {\sl Nat. Commun.} {\bf 4}, 1474 (2013).



\bibitem{book}  Haug, H. \& Koch, S. W. {\it Quantum theory of the
    optical and electronic properties of semiconductors}, Chap. 10,
  (World Scientific, 2009).

\bibitem{x2} Ramasubramaniam, A. Large excitonic effects in
    monolayers of molybdenum and tungsten dichalcogenides. {\sl Phys. Rev. B} {\bf 86}, 115409 (2012).



\bibitem{shen} Ugeda, M. M. {\sl et al.} Observation of giant
    bandgap renormalization and excitonic effects in a monolayer
    transition metal dichalcogenide semiconductor. arXiv:1404.2331,
  Nat. Mater. in press.





\bibitem{tau-x} Kumar, N., He, J., He, D., Wang, Y. \& Zhao, H. Valley
  and spin dynamics in MoSe$_2$ two-dimensional crystals. {\sl
    Nanoscale}, {\bf 6}, 12690 (2014).


\bibitem{zhao} Kumar, N. {\sl et al.} 
 Exciton-exciton annihilation in MoSe$_2$ monolayers.
 {\sl Phys. Rev. B} {\bf 89}, 125427 (2014).


\bibitem{wagner}  Mermin, N. D. \& Wagner, H.  Absence of
  ferromagnetism or antiferromagnetism in one- or two-dimensional
  isotropic Heisenberg models. {\sl Phys. Rev. Lett.} {\bf
    17}, 1133-1136 (1966).


\bibitem{tio2-fp} Bhattacharya, P. {\sl et al.} 
   Room temperature
   electrically injected polariton laser. {\sl Phys. Rev. Lett.} {\bf 112},
  236802 (2014).



\bibitem{LO-mose1} Horzum, S. {\sl et al.} 
 Phonon softening and
    direct to indirect band gap crossover in strained single-layer
    MoSe$_2$. {\sl Phys. Rev. B} {\bf 87}, 125415 (2013).

\bibitem{LO-mose2} Jin, Z., Li, X., Mullen, J. T. \& Kim,
  K. W. Intrinsic transport properties of electrons and holes in
  monolayer transition-metal dichalcogenides. {\sl Phys. Rev. B} {\bf
    90}, 045422 (2014).




\bibitem{noda} Ogawa, S., Imada, M., Yoshimoto, S., Okano, M. \& Noda,
  S.  Control of light emission by 3D photonic
  crystals. {\sl Science} {\bf 305}, 227-229 (2004)

\bibitem{noda6}
Noda, S., Fujita, M., \& Asano, T. Spontaneous-emission control by photonic crystals
and nanocavities. {\sl Nature Photon.} {\bf 1}, 449-458 (2007).



\bibitem{cdte} Jiang, J. H. \&  John, S.  Photonic crystal architecture for room temperature equilibrium
  Bose-Einstein condensation of exciton-polaritons. {\sl Phys. Rev. X} {\bf
  4}, 031025 (2014).

\bibitem{method3} Chamlagain, B. {\sl et al.}  {\sl ACS Nano} {\bf 8}, 5079-5088
(2014)

\bibitem{method2} Wang, X. {\sl et al.} Chemical vapor deposition
  growth of crystalline monolayer MoSe$_2$. {\sl ACS Nano} {\bf
  8}, 5125-5131 (2014).

\bibitem{cvd} Shim, G. W. {\sl et al.}  Large-area single-layer
  MoSe$_2$ and its van der Waals heterostructures. {\sl ACS Nano} {\bf 8},
  6655-6662 (2014).

\bibitem{yang1} John, S. \& Yang, S.  Electromagnetically induced
  exciton mobility in a photonic band gap. {\sl Phys. Rev. Lett.} {\bf 99},
  046801 (2007).

\bibitem{yang2}  Yang, S. \& John, S. Exciton dressing and capture
  by a photonic band edge. {\sl Phys. Rev. B} {\bf 75}, 235332 (2007).


 \bibitem{book-kavo} Kavokin, A. V. \& Malpuech, G. {\it Cavity polaritons}
   (Elsevier, Amsterdam, 2003), chap. 3.





\end{thebibliography}
\end{document}